\documentclass{aastex}
\usepackage{emulateapj5}
\usepackage{graphics}
\def\gtorder{\mathrel{\raise.3ex\hbox{$>$}\mkern-14mu
             \lower0.6ex\hbox{$\sim$}}}
\def\ltorder{\mathrel{\raise.3ex\hbox{$<$}\mkern-14mu
             \lower0.6ex\hbox{$\sim$}}}

\shorttitle{Gaussian Aperture Pupil Mask Observations}
\shortauthors{Debes, Ge \& Chakraborty}

\begin{document}
\title{First High Contrast Imaging Using a Gaussian Aperture Pupil Mask}
\author{John H. Debes\altaffilmark{1}, Jian Ge}
\author{and}
\author{Abhijit Chakraborty}
\affil{Pennsylvania State University}
\affil{Department of Astronomy \& Astrophysics, University Park, PA 16802, USA}
\altaffiltext{1}{NASA GSRP Fellow, debes@astro.psu.edu}

\begin{abstract}
Placing a pupil mask with a gaussian
aperture into the optical
 train of current telescopes represents a way to attain high contrast 
imaging that potentially improves contrast by orders of 
magnitude compared to current techniques.  We present here the first 
observations ever using a gaussian aperture pupil mask (GAPM)
on the Penn State near-IR Imager and Spectrograph (PIRIS) at the Mt. Wilson 
100$^{\prime\prime}$ telescope.  Two nearby stars were observed, $\epsilon$ 
Eridani and $\mu$ Her A.  A faint companion was detected 
around $\mu$ Her A, confirming it as a proper motion companion. Furthermore, 
the
observed H and K magnitudes of the companion were used to constrain its nature.
No companions or faint structure
 were observed for $\epsilon$ Eridani.  We found that our observations with the
 GAPM achieved contrast levels
similar to our coronographic images, without blocking light from the central 
star.  The mask's performance also 
nearly 
reached sensitivities reported for other ground based adaptive optics 
coronographs and deep HST images, but did not reach theoretically predicted 
contrast levels.  We outline ways that could improve the performance
of the GAPM by an order of magnitude or more.  
\end{abstract}

\section{Introduction}
The search to directly image an extrasolar Jovian planet requires contrast 
levels of $\sim$10$^{-9}$ a few $\lambda / D$ from the central star.  
Scattered light in a telescope
and the diffraction pattern of the telescope's aperture limit the contrast 
possible for direct detection of faint companions.  The circular aperture of 
telescopes creates a 
sub-optimal diffraction pattern, the so-called Airy Pattern which is 
azimuthally symmetric.  In addition, the intensity in the diffraction pattern 
of the circular
aperture declines as $\theta^{-3}$.  Currently the best way to diminish
the Airy pattern is to use a coronograph by using the 
combination of a stop in the focal plane that rejects a majority of the
central bright object's light and a Lyot stop in the pupil plane to reject high
frequency light \citep{lyot,malbet96,sivar01}.  Several recent ideas explore the use of alternative ``apodized'' apertures for high contrast
imaging in the optical or near-infrared \citep{nisenson01,spergel01, ge02}.
These ideas revisit ideas first experimented with in the field of optics 
\citep[][and references therein]{jacquinot64}.  

Apodizing an aperture rather than creating new technology to null or block a 
central bright object's light has several advantages.  Alignment of an 
object with an apodized aperture is not necessary, whereas for nulling 
interferometry or coronography it is essential.  The cost of creating a shaped
aperture is also quite low, if one makes the apodized aperture shape as a
pupil mask in the optical train.  This allows any current telescope to 
easily perform high contrast imaging.  Each telescope would then take 
images where the contrast would be only limited by atmospheric turbulence and
scattered light in the optical system.  

A promising design for a shaped aperture was recently suggested
by \citet{spergel01}.  In this case the top and bottom edges of the aperture
are described by gaussian functions:
\begin{eqnarray}
y_t & = & a R\left\{ \exp \left[-\left(\alpha x/R\right)^{2}\right] -\exp \left(-\alpha^{2}\right) \right\} \\
y_b & = & -b R\left\{\exp\left[-\left(\alpha x/R\right)^{2}\right] -\exp\left(-\alpha^{2}\right) \right\} 
\end{eqnarray}
where $x$ goes from $-R < x < R$.  The fourier transform of the aperture 
shape function gives the resulting 
diffraction pattern in the imaging plane.  Since the fourier transform of a 
gaussian is another gaussian, the intensity of the diffraction pattern along 
one image plane axis decreases exponentially, which we denote the high contrast
 axis.  The variables $a, b,$ and $\alpha$ are all free parameters that can be
used to optimize the aperture for depth of contrast, the angle from the central
object at which high contrast starts, and the azimuthal area of high contrast.
  
\begin{figure}
\plotone{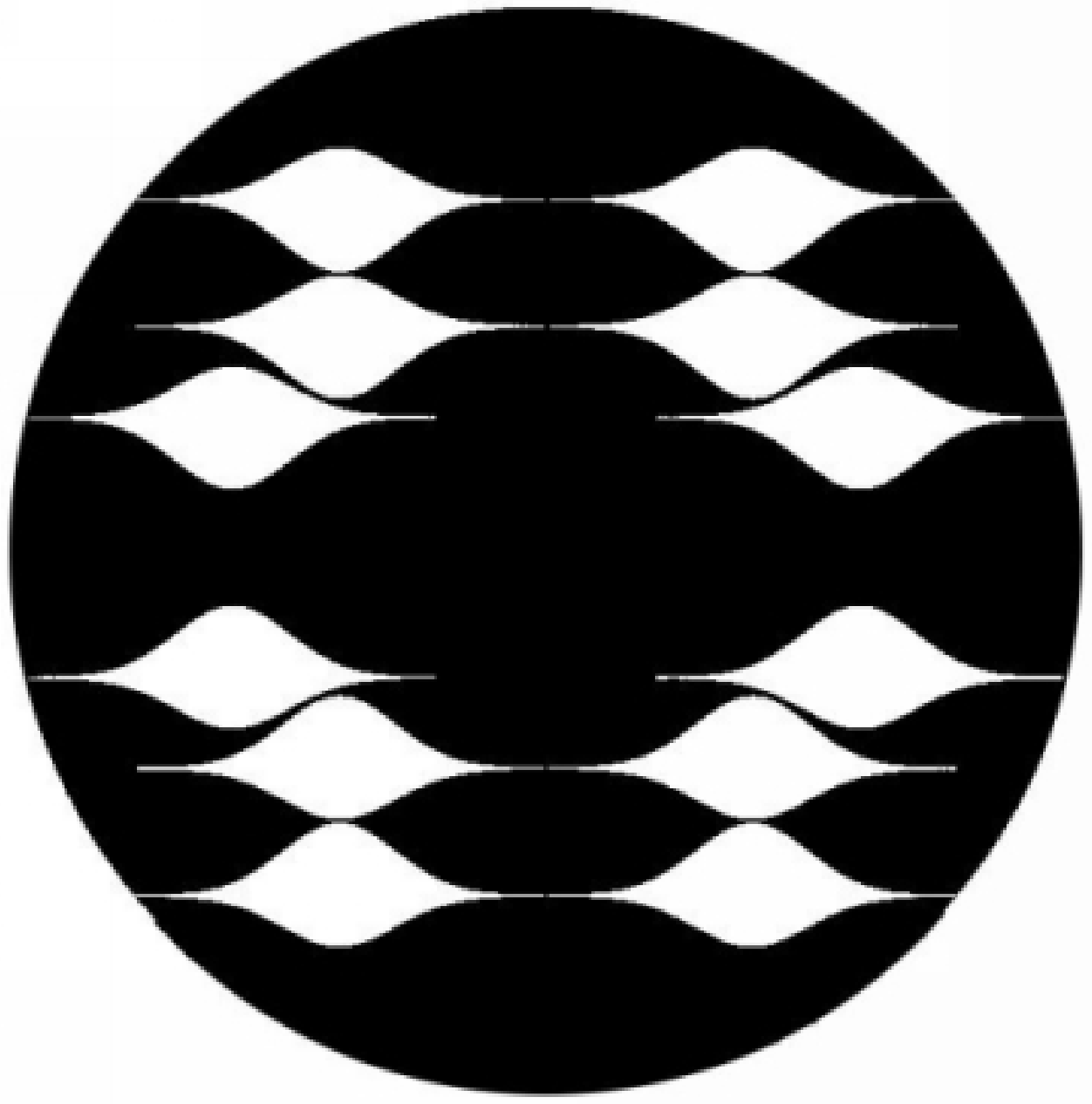}
\caption{\label{pupilfig} Design of the GAPM used at Mt. Wilson.  Twelve apertures were used
to provide throughput and avoid the support structure of the telescope.}
\end{figure}

In order to test the performance of a gaussian aperture pupil
mask on a telescope we 
designed a cold gaussian pupil mask and placed it in the optical train of the 
Penn State near-IR Imager and Spectrograph (PIRIS)  for
near-infrared imaging \citep{ge01,ge02}.  Figure \ref{pupilfig} shows our pupil mask
 design.  The placement of gaussian apertures avoided the support structure 
of the telescope to maximize contrast at the cost of
halving the resolution of the images.  Throughput was also an issue, so to 
achieve roughly 25\% throughput a total of 12 apertures were situated 
symmetrically
 in each quadrant of the pupil mask.  To maximize the azimuthal coverage so 
that in any one image roughly half of the azimuthal area was in a high contrast
region, we varied the height of the top and bottom of the apertures.  Full 
details of our design will be published in a future paper (Debes \& Ge 2002, in preparation).
The current pupil design would be well suited
for existing 10m telescopes to look for faint companions separated from their 
parent stars by 5 AU out to $\sim$40 pc in the visible or $\sim$10 pc for the 
near-IR limited only by atmospheric turbulence and scattered light in the 
telescope.  Scattered light is 
not a
hard limit; with a deformable mirror dedicated to correcting mirror 
surface errors, the levels of scattered light can be reduced by $\sim$1-2 
orders of magnitude \citep{malbet95}.

We report here the first results of observations using the GAPM to demonstrate
 the feasibility of using such a mask at an observatory.  While the
testing of optics with a GAPM has been mentioned before \citep[][e.g.]{jacquinot64}, to our knowledge
these are the first astronomical observations using a GAPM for detecting faint
companions. 
In Section \ref{obs} we will describe the 
observations we have done to date,
 and in Section \ref{res} we present our results.  Finally, in Section 
\ref{dis} we discuss the implications of our results.

\section{\label{obs} Observations}
The stars $\epsilon$ Eridani (GJ 144 = HD 22049 = HR 1084, V=3.72, d=3.2 pc)
and $\mu$ Her A (GJ 695A = HD 161797 = HR 6623, V=3.41, d=8.4 pc) were observed
 with the GAPM.  
The star $\epsilon$ Eridani was observed with seven 4s integrations in the K band,
while $\mu$ Her A was observed with one 3s and one 10s integration in the H 
band.  Both objects were also observed in normal imaging modes.  The plate 
scale and orientation for PIRIS was determined by measuring
the positions of several stars in the Orion Nebula and comparing them to 2MASS
 data.  We find that the plate 
scale of PIRIS is $.082^{\prime\prime} \pm .001^{\prime\prime}$
pixel$^{-1}$. 

The GAPM was aligned 23$^{\circ}$ clockwise from North to line up 
with the support structure of the telescope.  Therefore, the axis of greatest
contrast would be oriented perpendicular to the alignment of the mask or along
a line running from the NW to the SE in an image.

For our coronographic modes we used a stop in the focal plane that has a 
gaussian transmission function \citep{nakajima94} with a fwhm of 
$\sim$1$^{\prime\prime}$ in an image. 
A Lyot stop was placed in the pupil
 plane whose dimensions were chosen for an optimal combination of throughput 
and contrast (Ge \& Debes 2002; Ftaclas 2001, private communication).

\section{\label{res} Results}
\subsection{$\epsilon$ Eridani}
\begin{figure}
\plotone{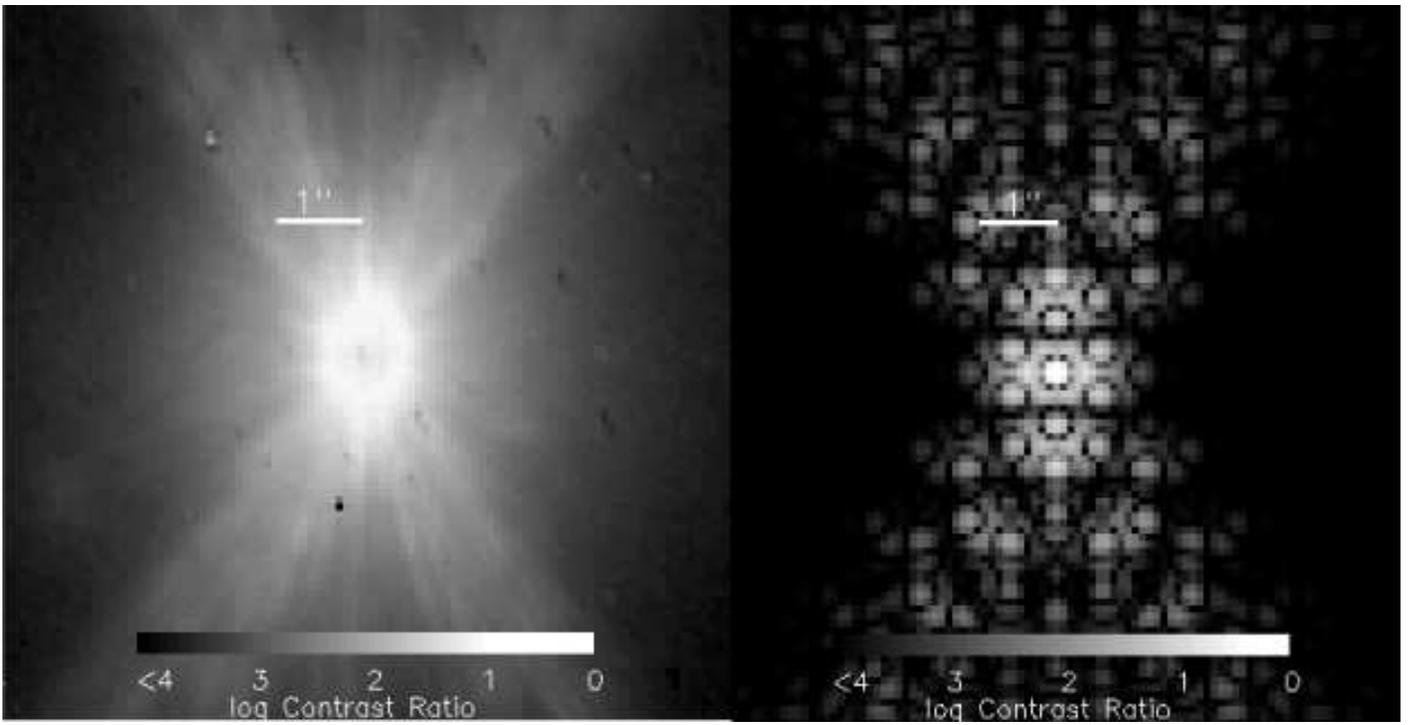}
\caption{\label{fig1} Co-added image of $\epsilon$ Eridani with a GAPM, 
the first observations ever done.  The right image shows a theoretically
produced PSF for comparison.    
The scale for the image is logarithmic to show high contrast features.}
\end{figure}

Figure \ref{fig1} shows a logarithmic scaled image of the PSF of the GAPM 
rotated so that the high contrast area is horizontal. 
The PSF of the GAPM is not azimuthally symmetric and the high contrast region
extends roughly over 180$^\circ$.  The majority of light in the diffraction 
pattern is spread 
into the four wings, leaving higher contrast regions 
between them.  For comparison, an image of the theoretical PSF of the GAPM 
is pictured.  No faint companions were detected, but fairly 
stringent limits can be imposed on what kind of companions would have been
detected between 1$^{\prime\prime}$ and 10$^{\prime\prime}$ from the central
star in the high contrast regions.

\begin{figure}
\plotone{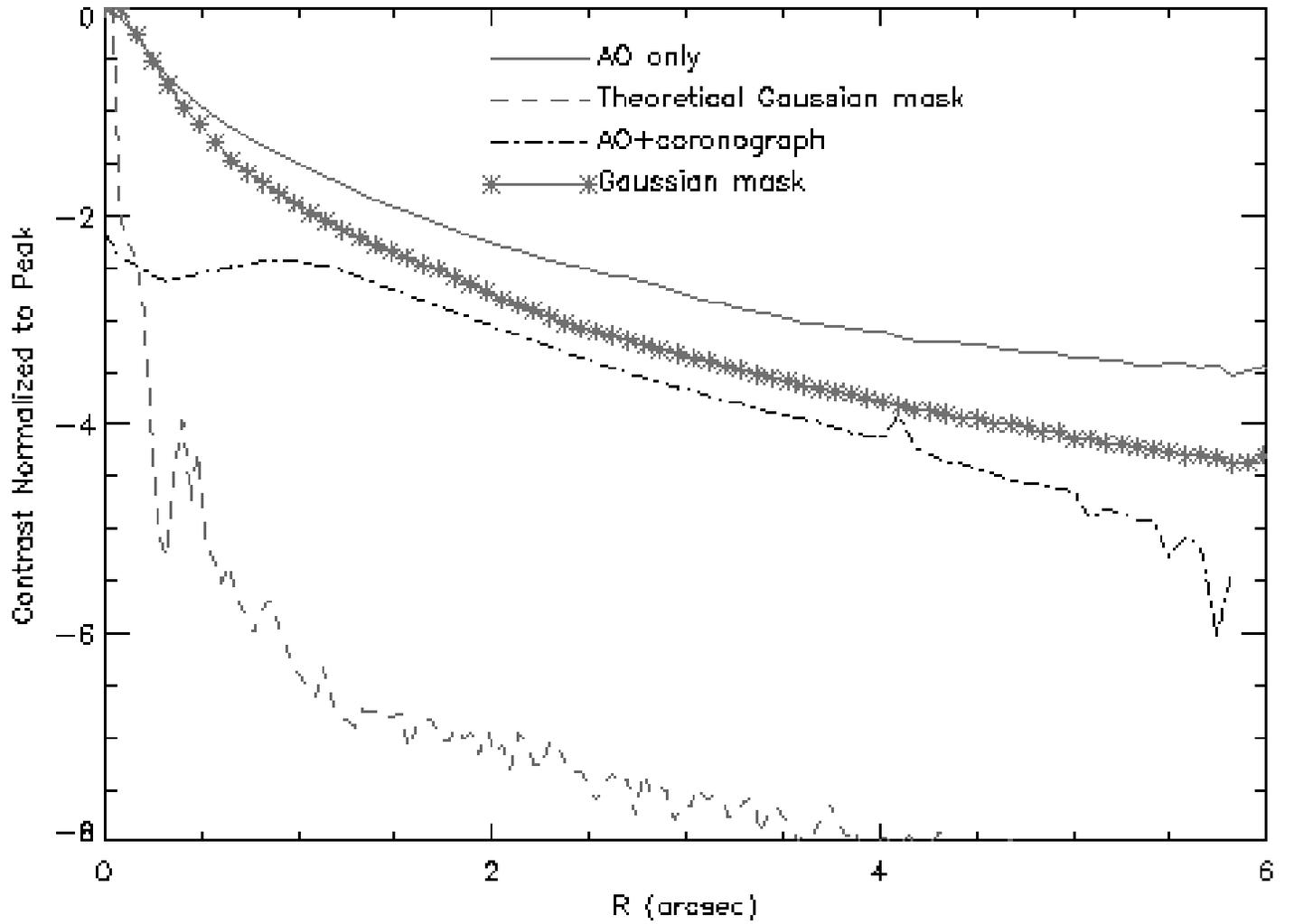}
\caption{\label{contrast} Comparisons of contrast achieved with the GAPM and 
other modes of PIRIS.  The GAPM performs only a factor of two worse than a
traditional coronograph without blocking any of the central star's light.}
\end{figure}

In order to test the level of contrast achieved by the gaussian mask in 
comparison to other modes of PIRIS, coronographic and adaptive optics 
observations were taken of the star+faint companion system Gl 105 \citep{gol95,
gol00}.  Figure
\ref{contrast} shows the differences in the three modes plus a simulation of
the performance of our pupil mask design with no scattered light or atmospheric
 turbulence.  The observations were azimuthally averaged in 
high contrast areas.
For Gl 105, we azimuthally averaged the PSF with the exception of 35$^\circ$ 
on either side of the faint companion.  The resulting
profiles were then normalized to peak flux.  Further scaling had to be
performed for the coronographic and gaussian pupil mask profiles.  In these two
cases the peak flux could not be determined due to the coronographic mask for 
Gl 105 and due to saturation for $\epsilon$ Eridani.  In the case of Gl 105 
this problem was circumvented by the presence of the companion which was used
to properly scale the flux.  The error in the curve for the coronograph is due
mostly to uncertainties in the positions  and flux ratio of the companion and the central star, but
should be no larger than on the order of 10\%.  
  In the case of $\epsilon$ Eridani we took 
unsaturated images of a dimmer standard star and normalized the azimuthally averaged PSF at .82$^{\prime\prime}$.  The profile of the GAPM is 
a composite of those two sets of observations.   
The error in this normalization is dominated by variations in the PSF and to estimate the effects of 
this we varied the point at which we normalized the two profiles.  We found that at most the error
introduced is of the order 25\%.

As can be seen in Figure \ref{contrast}, the GAPM
performs better than AO alone and is $\sim$2 times worse than a traditional
coronograph $>$ 1$^{\prime\prime}$.  This is without any attempt to
block the light of the central star.  We discuss avenues for improving the 
performance of the 
mask in Section \ref{dis}.  Comparision of normalized contrast versus
 $\lambda /D$
 to other coronographs and to deep HST WFPC2 observations of $\epsilon$ 
Eridani show that our gaussian pupil and coronographic mode performs similarly
 in terms
of contrast to previous work done \citep{schroeder00,hayward01,luhman02}.  
It is also evident
that several effects degrade contrast severely from the theoretical 
simulation.  A large part of this degradation is due to the scattered light
in the telescope and the AO.  We discuss the possible causes of contrast 
degradation in Section \ref{dis}.

From the azimuthally averaged profiles, limits to the type of companions we 
could have detected around $\epsilon$ Eridani can be estimated.  Beyond 4$^{\prime\prime}$ ($\sim$ 13 AU), any companion with $\Delta m_K$ $<$ 9.3 mag would
have been detected, and at 8$^{\prime\prime}$ ($\sim$ 26 AU) any companion 
$<$ 11.8 mag would have been detected.  These translate to $M_K$ of 13.6 and 
16.1 respectively.  Age is also an important consideration when talking about 
faint companions.  The age of $\epsilon$ Eridani is not well known, but should
 be $<$ 1 Gyr \citep{soderblom89,drake93}.  A more recent evaluation on the 
data for $\epsilon$ Eridani places the age at .73 $\pm$.2 Gyr \citep{song00}.  We looked at 
the models of \citet{burrows97} and \citet{chabrier00} to determine the range 
of possible companion masses that we could have detected based on .7 Gyr and 
M$_K$=13.6 and 16.1, corresponding to 38$\pm$8 M$_J$ and 20$\pm$5 M$_J$.  We 
derived these estimates by taking a value intermediate to the two models, 
while the error represents their spread.
  
\subsection{$\mu$ Her A}

\begin{figure}
\plotone{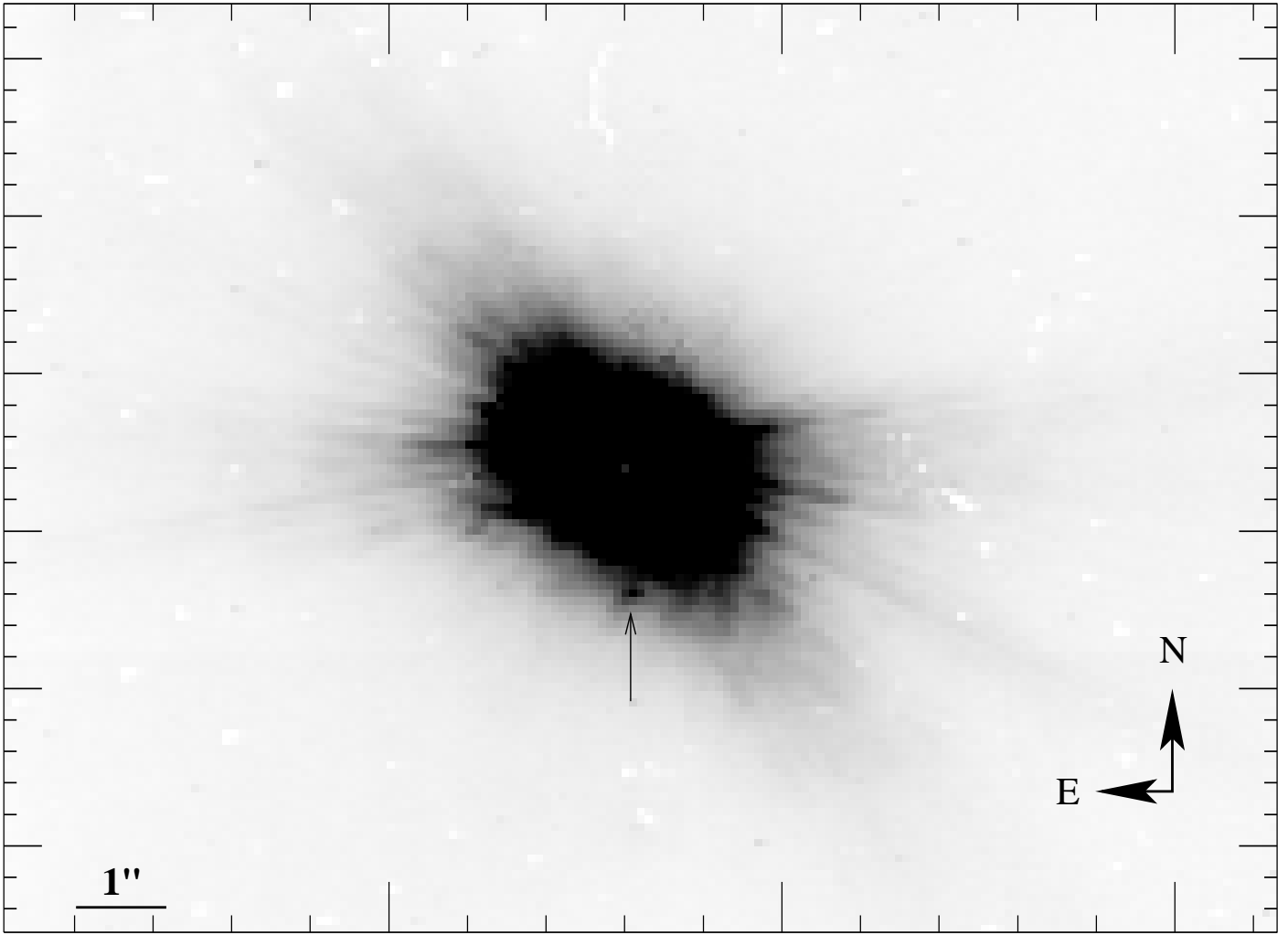}
\caption{\label{muher}Raw image of $\mu$ Her A and its faint companion taken with the GAPM.  The 
companion is clearly visible, and confirms it as a proper motion companion to $\mu$ Her A. 
}
\end{figure}

A faint companion around $\mu$ Her A was detected with high S/N in both 
GAPM and AO images.  Fig \ref{muher} shows the 
raw H image which clearly shows the faint companion.  
This companion was previously 
detected in R and I band AO images at Mt. Wilson, and used to explain a radial
 velocity acceleration corresponding to a $\sim$30 yr 
orbit \citep{turner01,cumming99,cochran94}.  Measurements of the
separation from AO images were complicated by the saturation of $\mu$ Her A, 
but the center
position was calculated by fitting a gaussian profile horizontally and 
vertically to unsaturated parts of the PSF.  We measured the separation of the
 companion 
to be 1.$^{\prime\prime}$3
$\pm$ $^{\prime\prime}$.2 ($\sim$ 11 AU) with a P.A. of 184$^\circ$ $\pm$ 8$^\circ$.  Errors 
represent 2-$\sigma$.  These measurements are consistent with those reported
by \citet{turner01}, although we measure a change in P.A. of $\sim$17$^\circ$.
The proper motion of $\mu$ Her A is -291.42 mas/yr in right ascension
and -750.03 mas/yr in declination, which implies that a much larger change
in separation would have been detected had the two objects not been bound \citep{perryman97}.  Our observations confirm
 the dimmer object to be a proper motion
companion and thus physically bound to the brighter star.

The nature of the companion has previously been suggested to be substellar 
from Monte Carlo calculations of probable mass based on observed angular
separation \citep{torres99}.  However, the observed photometry of the 
companion and 
the age of the primary star do not support that conclusion.  Unfortunately, 
standard star measurements with the GAPM could not be performed, but we 
took H and K photometry with the normal imaging mode of PIRIS.  We find that
 for the companion 
 $m_H$=8.5 $\pm$ .1 and $m_K$=7.9 $\pm$ .1, which corresponds to $M_H$=8.9 and
 M$_K$=8.3.  The age
of $\mu$ Her A has been estimated to be 10.6 Gyr by measurements of its 
magnetic activity \citep{barry87}.  However, there is a dependence on such 
measurements
based on the [Fe/H] of the star.  Using \citet{pinto98}'s correction for 
metallicity and [Fe/H]=.15 from \citet{strobel81} 
gives a corrected age of 4.2 Gyr.  \citet{turner01} reports a 
$\Delta m_{R}$=9.29 and $\Delta m_{I}$=7.26 between $\mu$ Her A and its
 faint companion.  Taking the observed $m_R$ and $m_I$ of the central star from \citet{jp11} one 
derives for the faint companion $m_R$=12.18 and $m_I$=9.77, corresponding to 
$M_R$=12.6 and $M_I$=10.2.  Along with the age, $M_R$, $M_I$, $M_H$, and $M_K$
 of the
 companion, we used the models of \citet{baraffe98,baraffe02} to conclude that
its mass should be $\sim$.13 M$_\odot$.  It should be noted that at this
 mass
the R and I band is anamolously bright by .4 and 1.2 mag respectively, while 
both H and K observations match the models to within the errors.  
\citet{baraffe98} mentions 
that the opacity around 1$\mu$m is not well known and when compared to
 observations, the models appear to overestimate the opacity in the R band, while underestimating opacity around the V band.  
In order to test this hypothesis \citet{baraffe98} arbitrarily increased the opacity in the 
V band and found an increase 
in flux in the I band.  At the same time this uncertainty had little
effect on the JHK bands.  We therefore regard the H and K magnitudes to be a 
more
reliable photometric indicator.  Even if the R and I magnitudes are a better
measurement to compare to models, the companion is above the hydrogen burning 
limit.  Spectra of the companion would be valuable in definitively settling 
its identity.  Such transition objects are extremely useful for constraining
low mass stellar models as well as determining the low mass end of the 
star formation rate.  
 
\section{\label{dis} Discussion}
These observations represent the first attempt at high contrast imaging with a
GAPM.  The relative ease and speed with which these first
masks have been produced and their performance presents a promising technology
that when mature may provide a viable alternative to coronography as a high 
contrast imaging mode.  The fact that this preliminary attempt achieves 
contrasts similar to modern techniques in coronography while not blocking any 
of the central star's light is significant and
encouraging for both ground based and space based attempts to directly image
extrasolar planets. Several factors have degraded the performance of the 
mask from reaching its theoretical sensitivities.

Our theoretical treatment of the mask neglects imperfect atmospheric 
correction, scattered light in the telescope, and the presence of a thermal 
background, all of which contribute to degrading the performance of the mask.
Images of the pupil plane (including the gaussian mask) show the presence of 
a small diffuse thermal background in the areas that have been blocked and 
should not emit light, on the order of 1\% the peak flux coming through the
gaussian openings.  Similarly, diffraction at the smallest openings of the 
gaussian aperture may wash out the gaussian shape and also significantly
degrade contrast.  Perhaps the most significant degradation comes from the
imperfect correction of the atmosphere in the AO system.  All of these factors
 must be quantified fully, which will be explored in a forthcoming paper 
(Debes \& Ge 2002, in preparation).

There are several ways to drastically improve the performance of the GAPM.
Suppression of any thermal background, widening the opening of the aperture to 
reduce diffraction effects, and higher precision construction of the masks all
 are
relatively quick ways to increase performance.  Additionally the GAPM should 
be able to be combined with a coronographic mask in the focal plane.  Based on
the performance of our coronographic modes that can suppress the light from the
central star by about an order of magnitude, we would expect our GAPM and
 coronographic mask to achieve sensitivities $\sim$ 5 times better than the 
coronographic mask and Lyot stop.  Longer term goals would be to design 
telescopes and instruments optimized for the suppression of excess scattered 
light and aberrations.

\acknowledgements

We thank D. McCarthy for loaning part of the optics for PIRIS, R. Brown for 
the PICNIC array, C. Ftaclas for coronographic masks, and A. Kutyrev for 
filters.  Several important discussions with D. Spergel were crucial in 
our understanding of gaussian apertures.  
We would also 
like to thank the invaluable help of
the Mt. Wilson staff and L. Engel for design help with the GAPM.

J.D acknowledges funding by a NASA GSRP fellowship under grant NGT5-119. This 
work was supported by NASA with grants NAG5-10617, NAG5-11427 as well as the 
Penn State Eberly College of Science.
\bibliographystyle{plainnat}

\clearpage

\end{document}